\documentclass[conference]{IEEEtran}
\IEEEoverridecommandlockouts
\usepackage{cite}
\usepackage{amsmath,amssymb,amsfonts}
\usepackage{newtxmath}
\usepackage{algorithmic}
\usepackage{graphicx}
\usepackage{booktabs}
\usepackage{textcomp}
\usepackage{xcolor}
\usepackage{soul}
\def\BibTeX{{\rm B\kern-.05em{\sc i\kern-.025em b}\kern-.08em
    T\kern-.1667em\lower.7ex\hbox{E}\kern-.125emX}}
\usepackage{graphicx}

\usepackage[hyphens]{url}
\usepackage{breqn}

\usepackage{caption}
\usepackage{subcaption}

\begin{document}

\title{A Hybrid Antenna Switching Scheme for Dynamic Channel Sounding
\thanks{This work has been funded by the Horizon Europe EU Framework Programme under the Marie Skłodowska-Curie grant agreement No.\,101059091, the Horizon 2020 EU Framework Programme under Grant Agreement No.\,861222, the Swedish Research Council (Grant No. 2022-04691), the strategic research area ELLIIT, Excellence Center at Linköping – Lund in Information Technology, and Ericsson.
}
}

\author{
\IEEEauthorblockN{$\text{Ali Al-Ameri}^{*}$, $\text{Jaeyoung Park}^{*}$,  $\text{Juan Sanchez}^{*}$, Xuesong Cai, and Fredrik Tufvesson}
\IEEEauthorblockA{Department of Electrical and Information Technology, \textit{Lund University}, Lund, Sweden \\
\{aalameri549, jpark9396\}@gmail.com \{juan.sanchez, xuesong.cai, fredrik.tufvesson\}@eit.lth.se \\
\footnotesize \textsuperscript{*}These authors contributed equally as first authors, and the order of names in the list is solely alphabetical.}



}

\maketitle

\begin{abstract}
Channel sounding is essential for the development of radio systems. One flexible strategy is the switched-array-based channel sounding, where antenna elements are activated at different time instants to measure the channel spatial characteristics. Although its hardware complexity is decreased due to fewer radio-frequency (RF) chains, sequentially switching the antenna elements can result in aliasing in the joint estimation of angles and Doppler frequencies of multipath components (MPCs). Therefore, pseudo-random switching has been proposed to mitigate such aliasing and increase estimation accuracy in both angular and Doppler domains. Nevertheless, the increased Doppler resolution could cause additional post-processing complexity of parameter estimation, which is relevant when the Doppler frequencies are not of interest, e.g., for spatial channel modeling. This paper proposes an improved hybrid sequential and random switching scheme. The primary purpose is to maintain the estimation accuracy of angles of MPCs while decreasing the resolution of Doppler frequencies for minimized complexity of channel parameter estimation. A simulated-annealing algorithm is exploited to obtain an optimized switching sequence. The effectiveness of the proposed scheme is also demonstrated with a realistic antenna array.
\end{abstract}

\section{Introduction} \label{sec:introduction}

Channel sounding is a fundamental task for designing and evaluating wireless systems \cite{THzMagazine, molisch2010wireless}. For modern wireless systems with multiple, and possibly a massive number of, transmit (Tx) and receive (Rx) antennas, it is crucial to understand the dynamic double-directional dual-polarized multiple-input multiple-output (MIMO) channels. One way is to exploit the so-called switched arrays, where a single RF chain is connected to multiple antenna elements via switches at the Tx/Rx sides \cite{bdb4a4ed338f46c8ae72fc4972961c13}. By activating different antenna pairs, channels of all antenna pairs are measured at different time instants. Switched-array sounding can accomplish one MIMO channel snapshot\footnote{A snapshot contains the channels of all antenna combinations.} within a very short measurement time, which is a substantial advantage for measuring dynamic MIMO channels compared to using, e.g., mechanically virtual antenna arrays \cite{xuesong3, 8901446, xuesong2}. These characteristics allow this approach to offer a good trade-off between cost and performance, elevating it as a popular sounding approach used nowadays.

The switching scheme, i.e.,  the order of activated antenna pairs, directly affects the channel parameter estimation. For example, if the antenna array (e.g., a uniform linear array (ULA)) has a regular geometry and the antenna elements are switched sequentially, the angle of a multipath component (MPC) can lead to linear phase variation of the received signals over time. Since the Doppler frequency also introduces linear phase variation, there could be multiple solutions of Doppler frequency and angle that result in the same received signals. This means that aliasing can arise when estimating angles of arrival/departure and Doppler frequencies of MPCs from the measurement data. To overcome this problem, a non-sequential (or pseudo-random) switching scheme was proposed in \cite{1258621}. The authors demonstrated the limitation of sequential antenna switching and the benefits of non-sequential antenna switching. However, the demonstration was conducted using arrays with regular geometry and omni-directional radiation patterns. In \cite{8728188}, the authors treated the switching sequence design as an optimization problem and proposed a simulated-annealing algorithm to efficiently obtain an optimized switching sequence to overcome aliasing. Antenna arrays with arbitrary geometries and realistic radiation patterns can be considered. The switching scheme finally obtained is fully scrambled. The effective measurement time of an MPC is highly likely to be the snapshot time, which results in a high Doppler resolution.

However, there are situations where the primary focus is on estimating propagation angles, e.g., for spatial channel modeling, and a highly accurate Doppler frequency estimation is not of interest. \textit{In these situations, the extra estimation complexity of the Doppler frequency, due to finer grid search in the initialization and/or final estimation steps, is a waste of computational time and memory.} Instead, a reduction of the Doppler frequency estimation accuracy of an MPC while maintaining the angular estimation accuracy would be preferred. To the best of the authors' knowledge, this opportunity has not been addressed in the literature, and the investigation presented in \cite{8728188} has been the most advanced work on switching sequence design in switched-array channel sounding.

To this end, this paper presents an improved simulated-annealing algorithm to obtain a hybrid sequential and random switching scheme. 
The hybrid switching scheme can reduce the resolution of Doppler frequencies and consequently the overall computation complexity of channel parameter estimation, with maintained angular accuracy. The algorithm's performance is analyzed theoretically and verified via simulations with realistic antenna array radiation patterns at millimeter-wave (mmWave) frequencies. 

The rest of the paper is structured as follows. Sect.\,\ref{sec:modelAndCRLB} introduces a general signal model for wireless channels and the estimation bounds on angles and Doppler frequency. 
Sect.\,\ref{sec:hybridScheme} presents the simulated-annealing algorithm to achieve the optimized hybrid switching scheme. 
In Sect.\,\ref{sec:hybridVerification}, simulation results verifying the scheme's performance are shown. Conclusive remarks and future work scope are finally included in Sect.\,\ref{sec:conclusions}.



The symbol notation throughout this paper follows: Bold upper case letters, e.g., $\mathbf{B}$, denote matrices. Bold lower case letters, e.g., $\mathbf{b}$, denote column vectors. $[\mathbf{B}]_{i,j}$ denotes the element in the $i$-th row and $j$-th column of the matrix $\mathbf{B}$ while $[\mathbf{b}]_{i}$ denotes the $i$-th element of the vector $\mathbf{b}$. Superscripts $\textsuperscript{T}$ and $\textsuperscript{H}$ denote transpose and Hermitian transpose. The operators $\otimes$ and $\odot$ denote Kronecker and Hadamard product. The operators $|\mathbf{\cdot}|$ and $||\mathbf{\cdot}||$ denote the Absolute-value norm and Euclidean norm. 

\section{Signal model and estimation bounds}
\label{sec:modelAndCRLB}



This section introduces a general signal model for a switched-array channel sounder. The error bounds of parameter estimation for angles and Doppler frequency are also derived in a SIMO scenario to demonstrate the necessity of hybrid switching.

\subsection{Signal Model}

Let us consider a case with $M_{\text{T}}$ Tx antennas, $M_{\text{R}}$ Rx antennas, and $M_t$ MIMO snapshots taken. We assume that the measurement time of $M_t$ MIMO snapshots is smaller than the coherence time of the channel\footnote{The maximum numnber of snapshots $M_t$ is determined by the coherence time of the channel. Coherence means that the structural parameters (angles, Doppler frequencies, delays, etc.) of MPCs are constant.}, and that the antenna array responses are constant within the measurement bandwidth with $M_f$ frequency points. The general vectorized data model for $P$ MPCs is given by
\begin{equation} \label{eq:general_vectorized_data_model}
  \begin{aligned}
    \mathbf{s}(\boldsymbol{\theta}_{\text{sp}})=  
    \sum_{p=1}^{P} \mathbf{B}(\boldsymbol{\mu}_{p})\cdot\boldsymbol{\gamma}_p,
 \end{aligned}
\end{equation}
where $\boldsymbol{\mu}_p$ includes the structural parameters for the $p$-th path, $\boldsymbol{\gamma}_p \in \mathbb{C}^{4}$ contains the polarimetric transmission coefficients for the $p$-th path, $\boldsymbol{\theta}_{\text{sp}} = \{ \boldsymbol{\mu}_p, \boldsymbol{\gamma}_p: \, p = 1,\dots,P \}$, and $\mathbf{B}(\boldsymbol{\mu}_p) \in \mathbb{C}^{M_t M_{\text{T}} M_{\text{R}} M_f \times 4}$ is the basis matrix for a single path. 
Further details of the data model can be found in \cite{8019844}, \cite{article}. For dynamic channels, we have
\begin{equation} \label{eq:generic_basis_matrix}
 \begin{aligned}
    \mathbf{B}(\boldsymbol{\mu}_{p}) =
        {\begin{bmatrix}
        (((\bold{b}_t \otimes \bold{b}_{\text{T}_{\text{H}}} \otimes \bold{b}_{\text{R}_{\text{H}}}) \odot \bold{a}_{\nu}) \otimes \bold{b}_f)^{\text{T}}\\
        (((\bold{b}_t \otimes \bold{b}_{\text{T}_{\text{H}}} \otimes \bold{b}_{\text{R}_{\text{V}}}) \odot \bold{a}_{\nu}) \otimes \bold{b}_f)^{\text{T}}\\
        (((\bold{b}_t \otimes \bold{b}_{\text{T}_{\text{V}}} \otimes \bold{b}_{\text{R}_{\text{H}}}) \odot \bold{a}_{\nu}) \otimes \bold{b}_f)^{\text{T}}\\
        (((\bold{b}_t \otimes \bold{b}_{\text{T}_{\text{V}}} \otimes \bold{b}_{\text{R}_{\text{V}}}) \odot \bold{a}_{\nu}) \otimes \bold{b}_f)^{\text{T}}\\
        \end{bmatrix}}^{\text{T}},
 \end{aligned}
\end{equation}
where $\bold{b}_t\in \mathbb{C}^{M_t}$ is the Doppler-induced change of responses due to the different starting time instants of MIMO snapshots, $\bold{b}_{\text{T}_{\text{H}}}, \bold{b}_{{\text{T}_{\text{V}}}}\in \mathbb{C}^{M_{\text{T}}}$ represent the polarimetric Tx array responses at the horizontal and the vertical polarizations, respectively,  $\bold{b}_{{\text{R}_{\text{H}}}}, \bold{b}_{{\text{R}_{\text{V}}}}\in \mathbb{C}^{M_{\text{R}}}$ represent the polarimetric Rx array responses, and $\bold{b}_f\in \mathbb{C}^{M_f}$ is the frequency basis vector depends on path delay. Lastly, $\bold{a}_{\nu}\in \mathbb{C}^{M_tM_{\text{T}}M_{\text{R}}}$ represents the Doppler-induced change of responses for different antenna pairs in every MIMO snapshot. 
Specifically, 
\begin{equation}
  \begin{aligned}\relax
        [&\bold{a}_{\nu}]_{m+(m_t-1)\cdot M_{\text{T}}M_{\text{R}}} = e^{j2\pi{\nu}_p[\boldsymbol{\eta}]_{m+(m_t-1)\cdot M_{\text{T}}M_{\text{R}}}}, \  \\ & m = 1, \dots, M_{\text{T}}M_{\text{R}}; \quad m_t = 1, \dots, M_{\text{t}},
   \end{aligned}
   \label{phase weighting vector}
\end{equation} where $\nu_p$ is the Doppler frequency of the $p$-th MPC, and $[\boldsymbol{\eta}]_{m+(m_t-1)\cdot M_{\text{T}}M_{\text{R}}}$ is the time instant when the $m$-th antenna pair is activated in the $m_t$-th snapshot relative to the starting time instant of the $m_t$-th snapshot. Note that the activation time instants can be independent of the snapshot index $m_t$, i.e., $[\boldsymbol{\eta}]_{m+(m_t-1)\cdot M_{\text{T}}M_{\text{R}}} = [\boldsymbol{\eta}]_{m+(m'_t-1)\cdot M_{\text{T}}M_{\text{R}}}$, $\forall m'_t \neq m_t$.

\subsection{Estimation bounds on angle and Doppler}
\label{subsec:EstimationBounds}


Without loss of the essence, let us consider a narrowband single-path channel that is measured for one snapshot with a vertically polarized omni-directional Tx antenna and a vertically polarized Rx ULA consisting of omni-directional elements.
The received signal $\mathbf{y\in \mathbb{C}^{M_{\text{R}}}}$ can be written as
\begin{equation} \label{eq:crlb_signalmodel}
\mathbf{y} = \mathbf{s} (\boldsymbol{\theta}_{\text{sp}}) + \mathbf{n} = \gamma_{\text{vv}} (\mathbf{b}_{\text{R}_{\text{v}}} \odot \mathbf{a}_{\nu}) + \mathbf{n},
\end{equation}
where $\mathbf{n}\in \mathbb{C}^{M_{\text{R}}}$ denotes zero-mean i.i.d circular white Gaussian noise with covariance matrix $\mathbf{R}_{nn}=\sigma^2\mathbf{I}_{M_{\text{R}}\times M_{\text{R}}}$, $\gamma_{\text{vv}} = re^{j\psi}$ is the complex amplitude of the path, 
and $\mathbf{b}_{\text{R}_{\text{V}}}$ is dependent on the azimuth of arrival (AOA) $\varphi$. The vector $\mathbf{b}_{\text{R}_{\text{V}}}$ has the form
\begin{equation}
    \mathbf{b}_{\text{R}_{\text{V}}} = 
        {\begin{bmatrix}
        e^{j\left(\frac{M_{\mbox{\tiny \text{R}}}-1}{2}\mu^{(\varphi)}\right)}, \dots , e^{-j\left(\frac{M_{\mbox{\tiny \text{R}}}-1}{2}\mu^{(\varphi)}\right)}
        \end{bmatrix}}^{\text{T}},
\end{equation}
with $\mu^{(\varphi)} = 2 \pi \frac{\Delta d}{\lambda} \cos{\varphi}$ being the structural parameter for the AOA $\varphi$, $\Delta d$ the inter-element separation of the ULA, and $\lambda$ the wavelength. It is clear from (\ref{eq:crlb_signalmodel}) that $\mathbf{y} \sim \mathcal{N}(\mathbf{s}(\boldsymbol{\theta}_{\text{sp}}),\,\sigma^2\mathbf{I}_{M_{\text{R}}\times M_{\text{R}}})$, where $\boldsymbol{\theta}_{\text{sp}}$ is the vector containing the parameters to be estimated. Therefore, 
\begin{equation} \label{}
   \boldsymbol{\theta}_{\text{sp}} = \begin{bmatrix}\varphi &\nu &r & \psi\end{bmatrix}^{\text{T}}.
\end{equation}
The variances for any vector $\hat{\boldsymbol{\theta}}_{\text{sp}}$ of unbiased parameter estimators can be bounded by the Cramér–Rao lower bound (CRLB) as
\begin{equation} \label{}
\text{var}([\hat{\boldsymbol{\theta}}_{\text{sp}}]_i) \geq \text{CRLB}([\boldsymbol{\theta}_{\text{sp}}]_i).
\end{equation}
These bounds can be expressed as
\begin{equation} \label{eq:crlb} 
 \text{CRLB}([\boldsymbol{\theta}_{\text{sp}}]_{i})=\frac{1}{[\boldsymbol{F}(\boldsymbol{\theta}_{\text{sp}},\sigma^2\mathbf{I}_{M_{\text{R}}\times M_{\text{R}}})]_{ii}},
\end{equation}
where $\boldsymbol{F}$ is the Fisher information matrix (FIM). Note that equation (\ref{eq:crlb}) assumes that the contributions of the off-diagonal entries of the FIM are negligible. By properly designing/optimizing the switching sequence, e.g., as done in \cite{1258621}, the off-diagonal entries can be close to 0. 
To find an expression for the FIM, the so-called Jacobian matrix is introduced as follows 
\begin{equation} \label{eq:jacobian}
\boldsymbol{D}(\boldsymbol{\theta}_{\text{sp}}) = \frac{\partial}{\partial \boldsymbol{\theta}_{\text{sp}}} \mathbf{s}(\boldsymbol{\theta}_{\text{sp}}).
\end{equation}
The FIM is related to the Jacobian and the covariance matrix by
\begin{equation} \label{eq:fisher}
\mathbf{F}(\boldsymbol{\theta}_{\text{sp}},\sigma^2 \mathbf{I}_{M_{\text{R}}\times M_{\text{R}}})=\frac{2}{\sigma^2}\Re \{ \mathbf{D}(\boldsymbol{\theta}_{\text{sp}})^{\text{H}}\cdot \mathbf{D}(\boldsymbol{\theta}_{\text{sp}})  \}.
\end{equation}
Now, calculating the Jacobian matrix with Equation (\ref{eq:jacobian}), using it to find the FIM according to Equation (\ref{eq:fisher}), and plugging the diagonal elements of the FIM into Equation (\ref{eq:crlb}) gives the following estimation bounds for the simple ULA case
\begin{equation} \label{eq:crlbphi}
\text{CRLB}(\varphi)=\frac{\sigma^2\cdot 6}{r^2\cdot M_{\text{R}} \cdot\left(M^2_{\text{R}}-1\right)} \cdot \left(\frac{\lambda}{2 \pi \cdot \Delta d \cdot \sin(\varphi)}\right)^2,
\end{equation}
\begin{equation} \label{eq:crlbnu}
 \text{CRLB}(\nu)=\frac{1}{8}\cdot \left(\frac{\sigma}{r\pi ||\boldsymbol{\eta}_{ \text{R}}||}\right)^2.
\end{equation}
Equation (\ref{eq:crlbphi}) shows that, for a given inter-element separation, carrier wavelength and AOA, the accuracy on the estimation of $\varphi$ depends solely on the number of receive antenna elements $M_{\text{R}}$. 
Equation (\ref{eq:crlbnu}) shows that the estimation accuracy of $\nu$ is influenced by the switching sequence vector $\boldsymbol{\eta}_{\text{R}}$, which has the form 
\begin{equation}
\label{eq:switchingVectorCanonical}
    \boldsymbol{\eta}_{\text{R}} = \mathbf{m}_{ \text{R}} \mathbf{P}_\pi \cdot \Delta t,
\end{equation}
where $\mathbf{m}_{\text{R}} = [-\frac{M_{\text{R}}-1}{2}, \dots , \frac{M_{\text{R}}-1}{2}]$, $\mathbf{P}_\pi \in \mathbb{N}^{M_{\text{R}} \times M_{\text{R}}}$ is any permutation matrix, and $\Delta t$ is the time difference between activating two neighboring antennas. It is clear that 
the longer/shorter the measurement time, the smaller/larger $\text{CRLB}(\nu)$ one can achieve, while it is possible to keep the same $\text{CRLB}(\varphi)$ for angle estimation. 

\section{Hybrid Switching Scheme} 
\label{sec:hybridScheme}




\subsection{Switching Sequence Optimization}
The simulated-annealing algorithm is known to find locally optimized switching sequences under certain constraints \cite{8728188,doi:10.1126/science.220.4598.671}. In this paper, we resort to the spatio-temporal ambiguity function \cite{8728188} as a core analysis tool, which is an extension of the ambiguity function presented in \cite{1143830}, \cite{723814}. The considered spatio-temporal ambiguity function has the form
\begin{equation}
    \label{eq:ambiguityCanonical}
    \begin{aligned}
    X(\boldsymbol{\mu}_{p},\boldsymbol{\mu}'_{p}, \boldsymbol{\eta}) & = \frac
    {\mathbf{b}^{\text{H}}(\boldsymbol{\mu}_{p}, \boldsymbol{\eta})\mathbf{b}(\boldsymbol{\mu}'_{p}, \boldsymbol{\eta})}
    {||\mathbf{b}^{\text{H}}(\boldsymbol{\mu}_{p}, \boldsymbol{\eta})|| \cdot ||\mathbf{b}(\boldsymbol{\mu}'_{p}, \boldsymbol{\eta})||},
    \end{aligned}
\end{equation}
where the vector $\mathbf{b}$ is the simplification of the general basis matrix $\mathbf{B}$ from (\ref{eq:general_vectorized_data_model}) when considering a single polarization. Notice the dependence on $\boldsymbol{\eta}$ explicitly highlighted from this section on, showing it as subject to the optimization procedure.


The objective function based on (\ref{eq:ambiguityCanonical}) can be expressed as
\begin{equation} \label{eq:objective_function}
    f_{\mathcal{P}}(\boldsymbol{\eta}) =  \iint_D |X({\boldsymbol{\mu}_p, \boldsymbol{\mu}'_p, \boldsymbol{\eta})}|^{\mathcal{P}} \,d\boldsymbol{\mu}_p\,d\boldsymbol{\mu}'_p,
\end{equation}
where $\mathcal{P}$ is the power of the objective function. We still consider a narrowband case, i.e., the worst case, where path delays cannot to used to distinguish MPCs. 
Then, the structural parameters $\boldsymbol{\mu}_p$ and $\boldsymbol{\mu}'_p$ include directions of departure, directions of arrival, and Doppler frequency under the following structure
\begin{equation} \label{}
  \begin{aligned}
\boldsymbol{\mu}_p=(\varphi_{\text{T}},\vartheta_{\text{T}},\varphi_{\text{R}},\vartheta_{\text{R}},\nu_p), \  \boldsymbol{\mu}'_p=(\varphi_{\text{T}}',\vartheta_{\text{T}}',\varphi_{\text{R}}',\vartheta_{\text{R}}',\nu'_p),
   \end{aligned}
\end{equation}
and the integration region $D$ is defined as
\begin{equation} \label{}
  \begin{aligned}
        D= \{\varphi_{\text{T}},\varphi_{\text{T}}', \varphi_{\text{R}},\varphi_{\text{R}}' \in[0,&\ 2\pi] \ \&   \\ \ \vartheta_{\text{T}}, \vartheta_{\text{T}}', \vartheta_{\text{R}}, \vartheta_{\text{R}}'\in[0,\ \pi]\ & \& \ \nu_p-\nu'_p\in[-\nu_{\text{up},p},\nu_{\text{up},p}] \}.
   \end{aligned}
\end{equation}

Algorithm 1 summarizes the steps to obtain an optimized switching sequence, given the objective function $f_{\mathcal{P}}(\boldsymbol{\eta})$, the so-called initial temperature $T_0$, the cooling rate $\alpha_0$, and the maximum number of iterations $k_{\text{max}}$. The update operation of the switching sequence, $\text{Update}(\boldsymbol{\eta})$, should be realized according to certain constraints, e.g., the existing random scheme by swapping two random elements in $\boldsymbol{\eta}$ or the improved hybrid scheme proposed in this work, which will be elaborated on in the following section. 
\begin{table}
\label{tab:algorithm}
\begin{tabular}{l}
\toprule
\textbf{Algorithm 1} The simulated-annealing algorithm to solve the optimization \\ problem \\ \midrule
1: Initialize $\boldsymbol{\eta},  \text{ the temperature } T=T_0,\text{ and } \alpha = \alpha_0$; \\
2: \textbf{while} $k \leq k_{\text{max}}$ \textbf{do}\\
3: \quad $\boldsymbol{\eta}' = \text{Update}(\boldsymbol{\eta})$ according to predefined constraints;\\
4: \quad $\textbf{if } \exp([(f_{\mathcal{P}}(\boldsymbol{\eta})-f_{\mathcal{P}}(\boldsymbol{\eta}')/T])>\text{random}(0,1) \textbf{ then}$\\
5: \qquad $\boldsymbol{\eta} = \boldsymbol{\eta'}$\\ 
6: \quad \textbf{end if}\\
7: \quad $T=\alpha T$\\
8: \textbf{end while}\\
\bottomrule
\end{tabular}
\end{table}



\subsection{Hybrid Switching Constraint}


From the $\text{CRLB}(\nu)$ derived in Section \ref{subsec:EstimationBounds}, the assumption of all antenna elements equally contributing to the estimation accuracy of channel parameters does not always hold. 
We can briefly modify the Rx antenna array to a version that includes unequal incident signal contributions at the different antenna elements, e.g., due to different radiation patterns of antenna elements and/or specific array geometries. 
A visually appealing example could be an assembly of four simple ULAs with patch antennas, shaped into a square array. The resulting square geometry ensures that only one side of the array is having a strong received signal upon the incidence of an MPC. In such cases, fully scrambled (random) switching is highly likely to result in the effective measurement time being the snapshot time, since the activated antenna goes back and forth between the effective side and the non-effective sides, which leads to a high Doppler resolution. 

Let us establish a threshold under which we can delimit subsets of antenna elements that receive significant signal contributions from a reduced range of directions of arrival. From there, it is possible to notice some subsets whose received signal contribution is negligible, and can be approximated to zero. Now, let us further constrain the switching sequence to be random within each subset, but sequential across subsets with the sequence following contiguous direction of arrival ranges. This proposed switching will be now referred to as $hybrid$ switching scheme. This means that we are constraining on the \textit{effective} measurement time, and are left with a group of \textit{effective} subsets of antenna elements $\tilde{\mathbf{m}}_{\text{R}} = [0, \dots , \tilde{M}_{\text{R}}-1]$ from which Doppler frequency estimation is carried out, where $\tilde{M}_{\text{R}} \leq M_{\text{R}}$. The hybrid scheme described here can be embedded into the permutation matrix $\mathbf{P}_\pi$ that composes the vector $\boldsymbol{\eta}_{\text{R,hyb}}$. Also, the approximation to the \textit{effective} received signal model yields a corresponding \textit{effective} permutation matrix $\tilde{\mathbf{P}}_\pi \in \mathbb{N}^{\tilde{M}_{\text{R}} \times \tilde{M}_{\text{R}}}$ that composes the vector $\tilde{\boldsymbol{\eta}}_{\text{R,rand}}$. If random switching is assumed for all the receive antenna elements when defining any other $\boldsymbol{\eta}_{\text{R,rand}}$, it is clear that $||\tilde{\boldsymbol{\eta}}_{\text{R,hyb}}||^2 = ||\tilde{\mathbf{m}}_{\mbox{\tiny \text{R}}} \tilde{\mathbf{P}}_{\pi} \cdot \Delta_t||^2 \leq ||\boldsymbol{\eta}_{\text{R,rand}}||^2$. This ultimately leads to $\text{CRLB}(\nu_{\text{hyb}}) \geq \text{CRLB}(\nu_{\text{rand}})$. The reduction in effective measurement time reduces the Doppler frequency estimation accuracy and thus the computational complexity and time in further estimation steps. Such reduction can be characterized by an effective factor $\xi$, which can be defined as
\begin{equation}
\label{eq:effectiveFactor}
    \xi = \frac{\tilde{M}_{\text{R}}}{M_{\text{R}}}.
\end{equation}

Coming back to the example of the four-sided square array, we could choose to regard each simple ULA as an antenna subset. Considering the geometrical implications and a total of $M_{\text{R}}$ antenna elements, we have $\tilde{M}_{\text{R}}=\frac{M_{\text{R}}}{4}$ effective antenna elements, and $\xi =\frac{1}{4}$. Since the effective antenna elements form one simple ULA, equation (\ref{eq:crlbphi}) holds for $\tilde{M}_{\text{R}}$, irrespective of the switching sequence. In contrast, optimized random and hybrid switching result in equivalent models of (\ref{eq:switchingVectorCanonical}) with time differences as $\widetilde{\Delta_t} \approx 4 \cdot \Delta_t$ and $\Delta_t$, respectively. Hence, hybrid switching reduces the effective measurement time, and consequently, the Doppler resolution with respect to random switching for the described square array.


Now, joining all the previous results into the proposed simulated-annealing algorithm, we can conceive a strategy to find an optimized hybrid switching sequence that reduces Doppler frequency estimation accuracy, while maintaining angular estimation accuracy and aliasing levels as for random switching. 
Algorithm 1 is used to find such a sequence, constraining it to abide to the hybrid switching scheme. 
The function $\text{Update}_{\text{hyb}}(\boldsymbol{\eta})$ swaps two random elements from one of the subsets at each iteration. The subset to be affected changes cyclically across iterations. In this way, random intra-subset switching and sequential inter-subset switching are guaranteed over the optimization procedure.

\section{Hybrid Switching Scheme Verification} \label{sec:hybridVerification}
\begin{figure}[h]
    \centering
    \includegraphics[width=0.4\columnwidth]{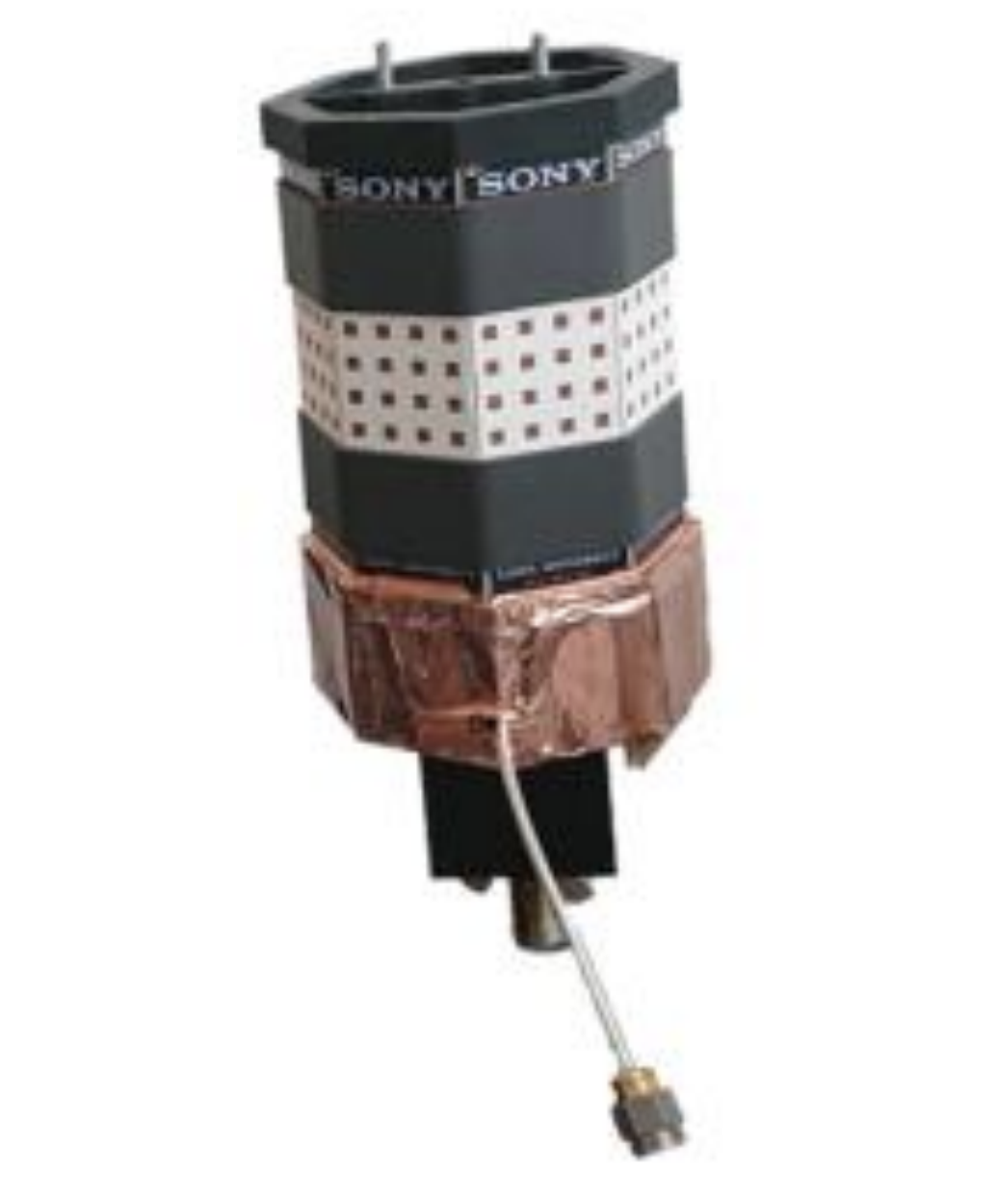}
    \caption{The octagonal Rx antenna array of the switched-array channel sounder at Lund University, Sweden.}
    \label{fig:mmWaveSounder}
\end{figure}

We use a realistic antenna array to verify the hybrid switching scheme via MATLAB simulations. For the Tx side, a single omni-directional Tx antenna is assumed. For the Rx side, the octagonal array of the switched array mmWave channel sounder \cite{bdb4a4ed338f46c8ae72fc4972961c13,9329997} at Lund University, Sweden, is exploited. As shown in Fig.\,\ref{fig:mmWaveSounder}, the octagonal array is composed of eight uniform rectangular array panels with 16 dual-polarized patch antennas arranged in a 4$\times$4 layout. The radiation patterns of all the $M_{\text{R}} = 128$ elements of the Rx octagonal array were measured in an anechoic chamber \cite{bdb4a4ed338f46c8ae72fc4972961c13,cai2023enhanced}, and the measured patterns are used in the simulations where we still consider the narrowband single-path single-polarization case.    


\subsection{Theoretical Expectations}



In the considered case, the basis matrix of the signal model can be simplified to the vector
\begin{equation}
\label{eq:basisSounder}
    \mathbf{b}(\boldsymbol{\mu}_{\text{R}}, \boldsymbol{\eta}_{\text{R}}) =
    {\begin{bmatrix}
      \bold{b}_{\text{R}_{\text{V}}} \odot \bold{a}_{\nu}
        \end{bmatrix}},
\end{equation}
where $ \boldsymbol{\mu}_{\text{R}}=(\varphi_{\text{R}},\vartheta_{\text{R}},\nu)$ is defined for a single MPC.
In turn, the ambiguity function further simplifies to 
\begin{equation}
\label{eq:ambiguitySounderSimplified}
    \begin{aligned}
        X_{\text{R}_{\text{V}}}(\boldsymbol{\mu}_{\text{R}},\boldsymbol{\mu}'_{\text{R}}, \boldsymbol{\eta}_{\text{R}}) = \frac
    {\mathbf{b}^{\text{H}}(\varphi_{\text{R}}, \vartheta_{\text{R}}, \nu, \boldsymbol{\eta}_{\text{R}})
    \mathbf{b}(\varphi'_{\text{R}}, \vartheta'_{\text{R}}, \nu', \boldsymbol{\eta}_{\text{R}})}
    {||\mathbf{b}^{\text{H}}(\varphi_{\text{R}}, \vartheta_{\text{R}}, \nu, \boldsymbol{\eta}_{\text{R}})|| \cdot ||\mathbf{b}(\varphi'_{\text{R}}, \vartheta'_{\text{R}}, \nu', \boldsymbol{\eta}_{\text{R}})||},
    \end{aligned}
\end{equation} and the objective function (15) applied in Algorithm 1 is modified accordingly. 
Two different update operations were performed, namely $\text{Update}_{\text{hyb}}(\boldsymbol{\eta})$ and
$\text{Update}_{\text{rand}}(\boldsymbol{\eta})$ to find out an optimized hybrid switching sequence and random switching sequence, respectively. The critical parameters in this algorithm followed the guidelines in \cite{8728188}, whereas $k_{\text{max}} = 200$.

Following the Rx array's geometry, it is natural to split the antenna elements into subsets of 16 antennas that correspond to each uniform rectangular array panel. It is geometrically expected that 3 of the 8 panels will receive the highest amount of power. This implies that the number of effective antenna elements receiving an MPC is around $\tilde{M}_{\text{R}} = 48$. Therefore, the expected effective factor with respect to an optimized random switching scheme is around $\xi = 3/8$.

\subsection{Simulation Results}
\begin{figure}[h]
     \centering
     \begin{subfigure}[b]{0.48\textwidth}
         \centering
         \includegraphics[width=\textwidth]{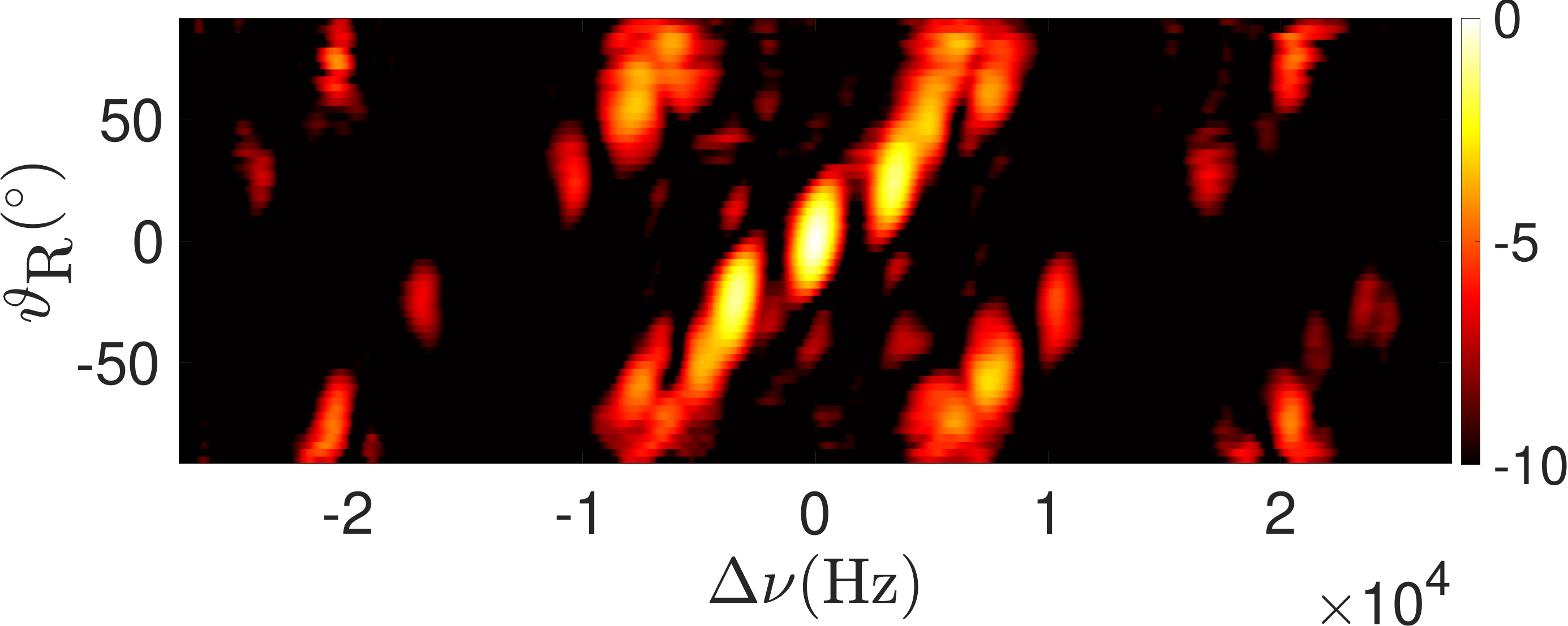}
         \caption{Under sequential switching}
         \label{fig:ambiguitySequentialEOA}
     \end{subfigure}
     \hfill
     \begin{subfigure}[b]{0.48\textwidth}
         \centering
         \includegraphics[width=\textwidth]{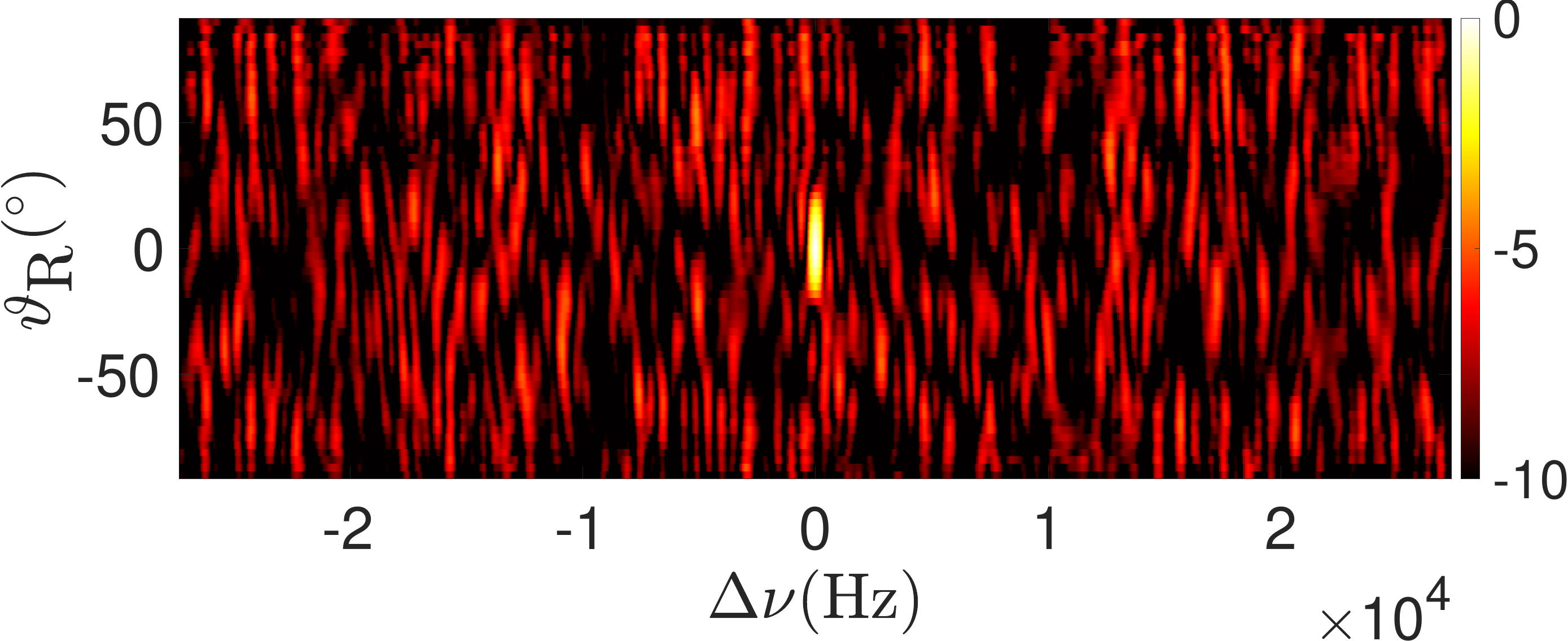}
         \caption{Under random switching}
         \label{fig:ambiguityRandomEOA}
     \end{subfigure}
     \hfill
     \begin{subfigure}[b]{0.48\textwidth}
         \centering
         \includegraphics[width=\textwidth]{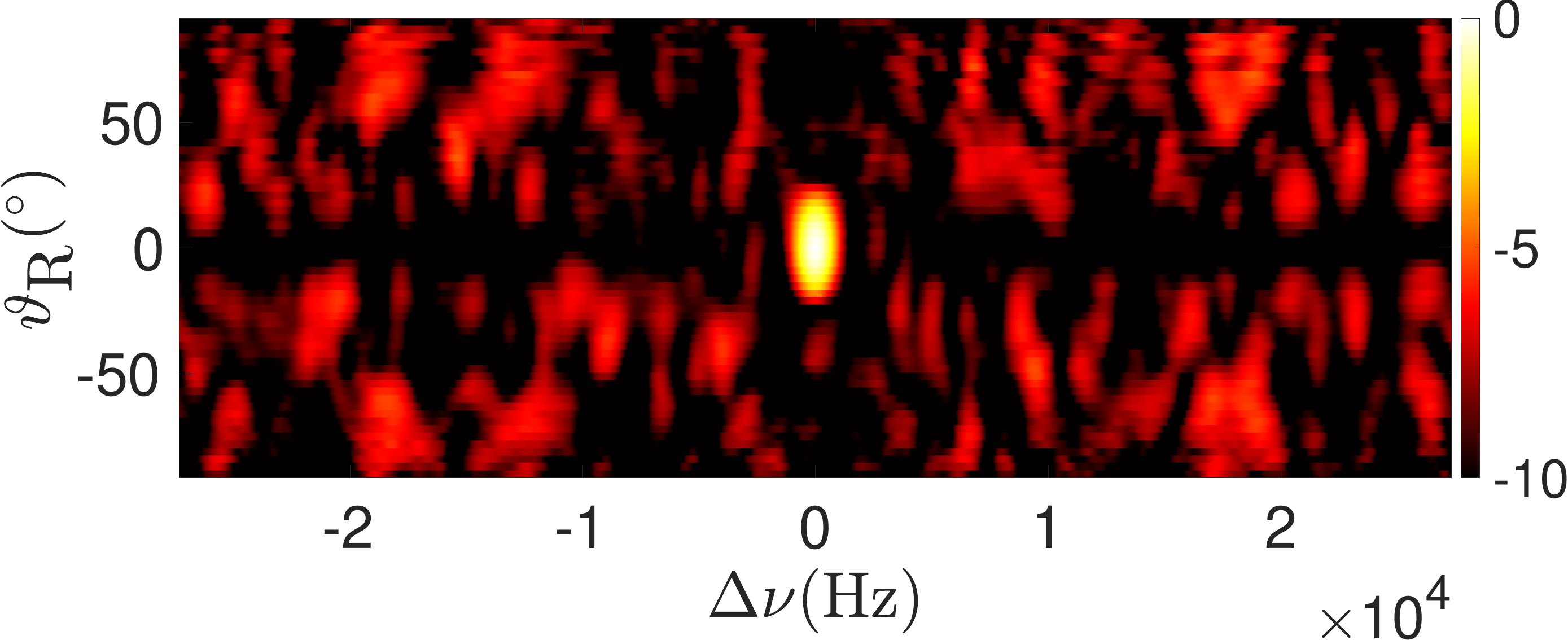}
         \caption{Under hybrid switching}
         \label{fig:ambiguityOptimizedEOA}
     \end{subfigure}
     \caption{Magnitude of optimized ambiguity function (in dB scale), $|X_{\text{R}_{\text{V}}}|$, for different switching schemes. X-axis: Doppler frequency; Y-axis: EOA} \label{fig:ambiguityEOA}
\end{figure}
After having run Algorithm 1 under both random and hybrid schemes for $k_{\text{max}} = 200$ iterations, with their respective update functions, the optimized switching sequences $\boldsymbol{\eta}_{\text{R}}^{\text{rand}}$ and $\boldsymbol{\eta}_{\text{R}}^{\text{hyb}}$ were found. The corresponding optimized ambiguity functions are plotted in Fig. \ref{fig:ambiguityRandomEOA} and Fig. \ref{fig:ambiguityOptimizedEOA}. Moreover, the ambiguity function of the sequential switching case is plotted in Fig. \ref{fig:ambiguitySequentialEOA}. 
As expected from \cite{1258621,8728188}, the level of aliasing is significantly reduced with respect to the sequential switching scheme when using non-sequential schemes, which can be seen by comparing Fig. \ref{fig:ambiguityRandomEOA} (or \ref{fig:ambiguityOptimizedEOA}) with Fig. \ref{fig:ambiguitySequentialEOA}.

Specifically, Fig. \ref{fig:ambiguityRandomEOA} shows the magnitude of the ambiguity function for $\boldsymbol{\eta}_{\text{R}}^{\text{rand}}$, in logarithmic scale, with respect to the elevation of arrival (EOA) and Doppler frequency, where $\Delta\nu = \nu - \nu'$.   
It can be observed that in Fig.\,\ref{fig:ambiguityRandomEOA} the half-power EOA range of $|X_{\text{R}_{\text{V}}}|$ lies around [$-16$, $18$]\,\textdegree{}, whereas the half-power Doppler frequency range lies around [$-273$, $273$]\,Hz. 
On the other hand, Fig. \ref{fig:ambiguityOptimizedEOA} shows the corresponding results under the hybrid scheme $\boldsymbol{\eta}_{\text{R}}^{\text{hyb}}$. Here, $|X_{\text{R}_{\text{V}}}|$ exhibits the same half-power EOA range as Fig.\,\ref{fig:ambiguityRandomEOA}. However, the half-power Doppler frequency range increases to around [-764, 764]\,Hz, which is 2.799 times broader than that of Fig. \ref{fig:ambiguityRandomEOA}. This increase approximately corresponds to the inverse of the effective factor $\frac{1}{\xi} = \frac{8}{3}$. With this, the theoretical results are verified by the simulation results. They both show that implementing an optimized hybrid switching scheme preserves the angular estimation accuracy and reduces the Doppler frequency accuracy with respect to an optimized random switching scheme. The reduction in Doppler frequency accuracy ultimately implies a reduction in the overall estimation complexity of the channel parameters. 

\begin{figure}[t]
    \centering
    \includegraphics[width=1\columnwidth]{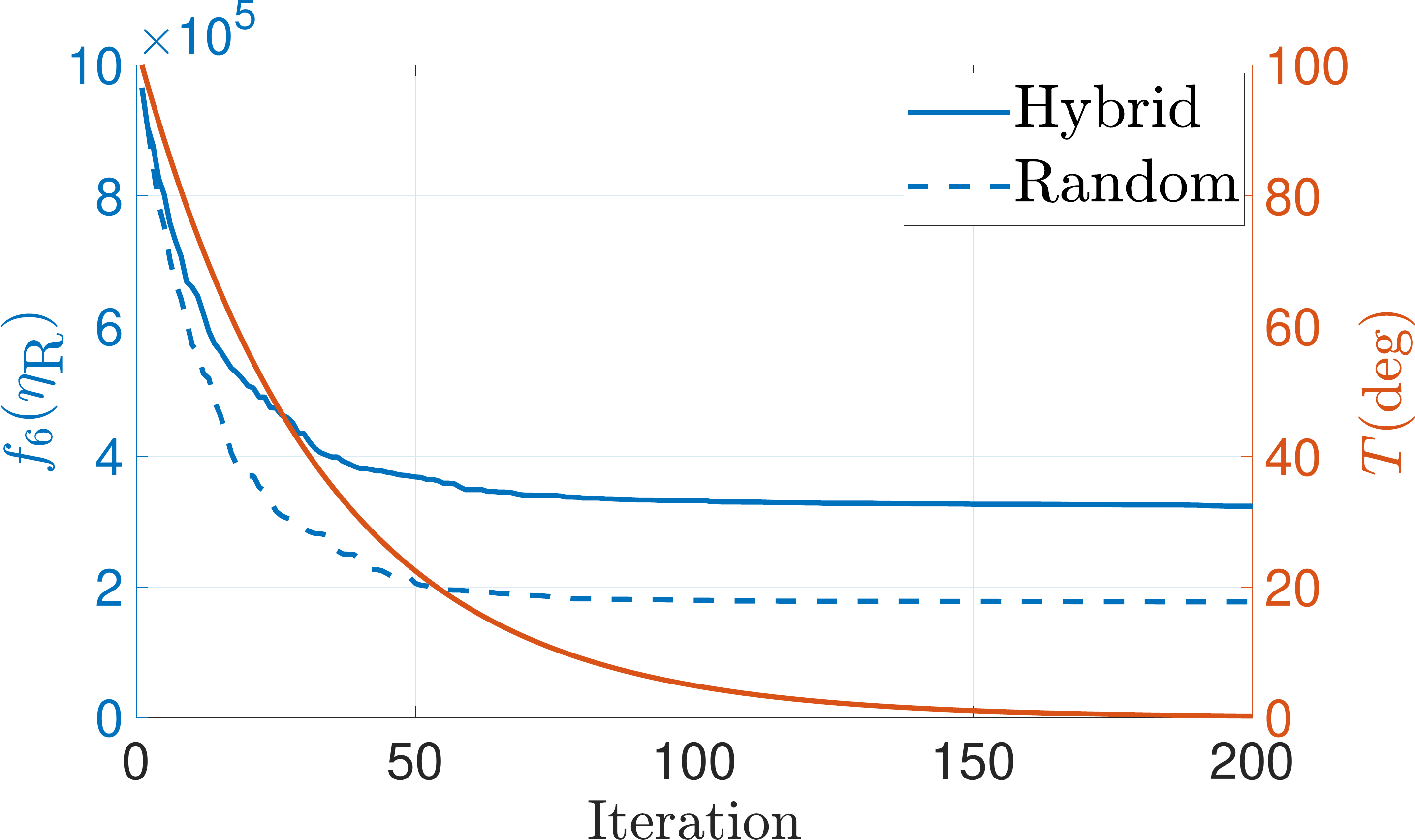}
     \caption{The evolution of the objective function and temperature during Algorithm 1.}
    \label{fig:EvolutionSA}
\end{figure}

In addition, Fig. \ref{fig:EvolutionSA} shows the evolution of the objective function $f_{6}(\boldsymbol{\eta}_{\text{R}})$ and the temperature $T$, across iterations of Algorithm 1, for the random and hybrid switching schemes. The vertical axis on the left edge shows the scale for both objective function curves, whereas the vertical axis on the right edge shows the scale for the temperature. Since the temperature decrease rate was the same for both schemes, the temperature curve remains the same. 
For the first iterations, the objective function for the random switching scheme decreases at a relatively higher rate than that of the hybrid switching scheme, also achieving a lower objective function value. Between iterations 50 and 100, when both curves stabilize, the decrease rates for both schemes converge to zero. The difference in the lower bounds of the objective functions for each scheme can be explained by the structural constraint in the hybrid scheme. The objective function measures the discrimination ability in the angular and Doppler domains in this case. 
Since the Doppler frequency estimation accuracy reduces under the hybrid scheme, its corresponding objective function inevitably converges asymptotically to a higher value than that of the random scheme.


\section{Conclusions} \label{sec:conclusions}
In this paper, we presented an improved hybrid switching scheme for switched-array-based channel sounding. 
This is realized by implementing a simulated-annealing-based optimization algorithm to randomize the switching within partitioned antenna element subsets while keeping sequential switching across these subsets. It can reduce the Doppler resolution and not compromise the estimation accuracy of angles. 
This was verified with realistic simulations using the measured radiation patterns of an octagonal antenna array from the mmWave channel sounder at Lund University. A panel-wise partitioning in this octagonal geometry allowed a maximum reduction in Doppler frequency estimation accuracy, as well as a preservation of the angular estimation accuracy when implementing a hybrid switching scheme for channel sounding. Future work will extend the investigation by considering the complete polarimetric channel model and further verify the estimation complexity and accuracy of the channel parameters with real channel measurements.


\bibliographystyle{IEEEtran}
\bibliography{ref}

\end{document}